\documentclass[twocolumn, aps, prd]{revtex4}
\usepackage{graphicx}

\newcommand{\Vec}[1]{\mbox{\boldmath$#1$}}

\begin{document}

\title{Electronic structure of an electron 
on the gyroid surface: A helical labyrinth}
\author{M. Koshino}
\affiliation{
Department of Physics, Tokyo Institute of Technology, 
Ookayama, Tokyo 152-8551, Japan}
\author{H. Aoki}
\affiliation{
Department of Physics, University of Tokyo, Hongo, Tokyo
113-0033, Japan}
\date{\today}

\begin{abstract}
Previously reported 
formulation for electrons on curved periodic surfaces is used to 
analyze the band structure of an electron bound on the gyroid surface
(the only triply-periodic minimal surface that has screw axes).  
We find that an effect of the helical structure appears as 
the bands multiply sticking together on the Brillouin zone boundaries.  
We elaborate how the band sticking is lifted when 
the helical and inversion symmetries of the structure are degraded.  
We find from this that the symmetries give rise to 
prominent peaks in the density of states.  
\end{abstract}

\maketitle



While the physics of electron systems in crystals
has been firmly established, it should be interesting to 
consider electrons bound on infinite, periodic {\it curved surfaces}, 
which we may envisage as crystals composed of surfaces.  
We have previously calculated the band structure of 
an electron confined on triply periodic (i.e., periodic along 
$x,y,z$) minimal surfaces (called P- and D-surfaces) and found that
bands and the Bloch wavefunctions 
are basically determined by the way in which the ``tubes'' 
are connected into a network.\cite{Koshino,KoshinoEP2DS}  

One asset of the crystal of surfaces is that 
we can deform them.  
We have indeed found that the P- and D surfaces, 
which are mutually Bonnet(conformal)-transformed, 
have related electronic structures.  
We can even twist the tubes, for which there is a special interest: 
Schoen pointed out in the late 1960's 
the following. P-surface (a cubic network of 
tubes) and D-surface (a diamond network) are typical 
triply periodic minimal surfaces, 
where minimal surfaces are defined as negatively-curved 
surfaces that have minimized areas with the mean
curvature ($\frac{1}{2}(\kappa_1+\kappa_2)$ with $\kappa_1,\kappa_2$
being the principal curvatures) vanishing everywhere on the
surface.  In addition, however, there exists a 
third one which Schoen called the gyroid(G-) surface.  
The P-, D-, and G-surfaces are related via the Bonnet
transformation, where the Bonnet angle (a parameter 
in the transformation) 
$\beta = 0, 38.0^{\circ}, 90^{\circ}$ correspond to P, G, D, 
respectively.  
G-surface is unique in its {\it triply helical} structure, 
on which there are no two-fold axes, nor straight lines, 
so we may call the surface a helical labyrinth.  
G-surface has been explored from 
the viewpoints of chemistry, crystallography and material science as well.
Namely, the gyroid structure is known to occur, topologically, in
some classes of crystal structures, which include 
clathrate compounds (such as Ba$_6$Ge$_{25}$\cite{Yamanaka,Yuan}), 
zeolite structures (such as MCM-48\cite{Carlsson}),
and ceramic structures fabricated with copolymer templates\cite{Chan}. 
A particular interest, naturally, is how the 
helical structure affects the electronic structure on G-surface.

Here we have calculated the band structure of
an electron bound on G-surface, 
adopting the formulation of our previous work\cite{Koshino}.
We find that an effect of the helical geometry 
appears as a set of {\it multiple band sticking} phenomena 
at the Brillouin zone edges.  
We have identified that the band sticking is related to 
the existence of screw axes and the inversion symmetry by 
checking that a degradation of 
the helical or inversion symmetries of the structure lifts 
the sticking, in an atomic model (``graphitic sponge'') 
realizing the structure.  We conclude 
from this that the symmetries give rise to 
prominent peaks in the density of states. 


We start by recapitulating the formulation for representing 
minimal surfaces. 
We consider a two-dimensional surface 
$\Vec{r}(u,v)  = (x(u,v),y(u,v),z(u,v))$
embedded in three spatial dimensions as parameterized by two coordinates $u,v$. 
When the surface is minimal, we can exploit 
the Weierstrass-Enneper representation given as
\begin{eqnarray}
  \Vec{r}(u, v) = {\rm Re} \left(
    \int^{w}_{w_0} (1 - w^2) F(w) {\rm \, d}w,  \right. \nonumber \\
   \left.
    \int^{w}_{w_0} i (1 + w^2) F(w) {\rm \, d}w,
    \int^{w}_{w_0} 2 w  F(w) {\rm \, d}w
  \right),
  \label{eqn:Weiermap}
\end{eqnarray}
where $w \equiv u+iv$ and 
$F(w) = ie^{i\beta}L/\sqrt{1 - 14w^4 + w^8}$ 
with the Bonnet angle $\beta = 38.015^{\circ}$ for G-surface,\cite{Terrones} 
and $L$ the linear dimension of the unit cell.  
As stressed in \cite{Koshino}, Weierstrass-Enneper representation 
that completely defines the surface is specified solely 
by its poles, which, in real space, correspond to 
`navels' (umbilical points) in differential-geometrical language, 
at which $\kappa_1=\kappa_2=0$.  
So a periodic minimal surface has negative curvatures everywhere 
except at the navels.

\begin{figure}
\begin{center}
\leavevmode\includegraphics[width=50mm]{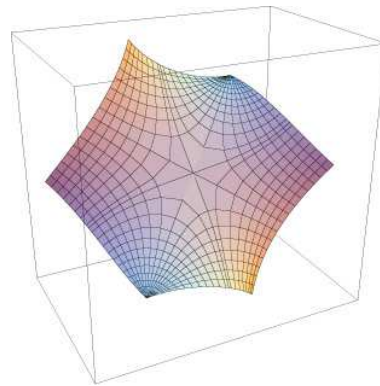}
\end{center}
\caption{
A primitive patch for G-surface.  Its center corresponds 
to the navel (see text), around which the surface is point-symmetric.  
}
\label{fig_patch}
\end{figure}

\begin{figure}
\begin{center}
\leavevmode\includegraphics[width=40mm]{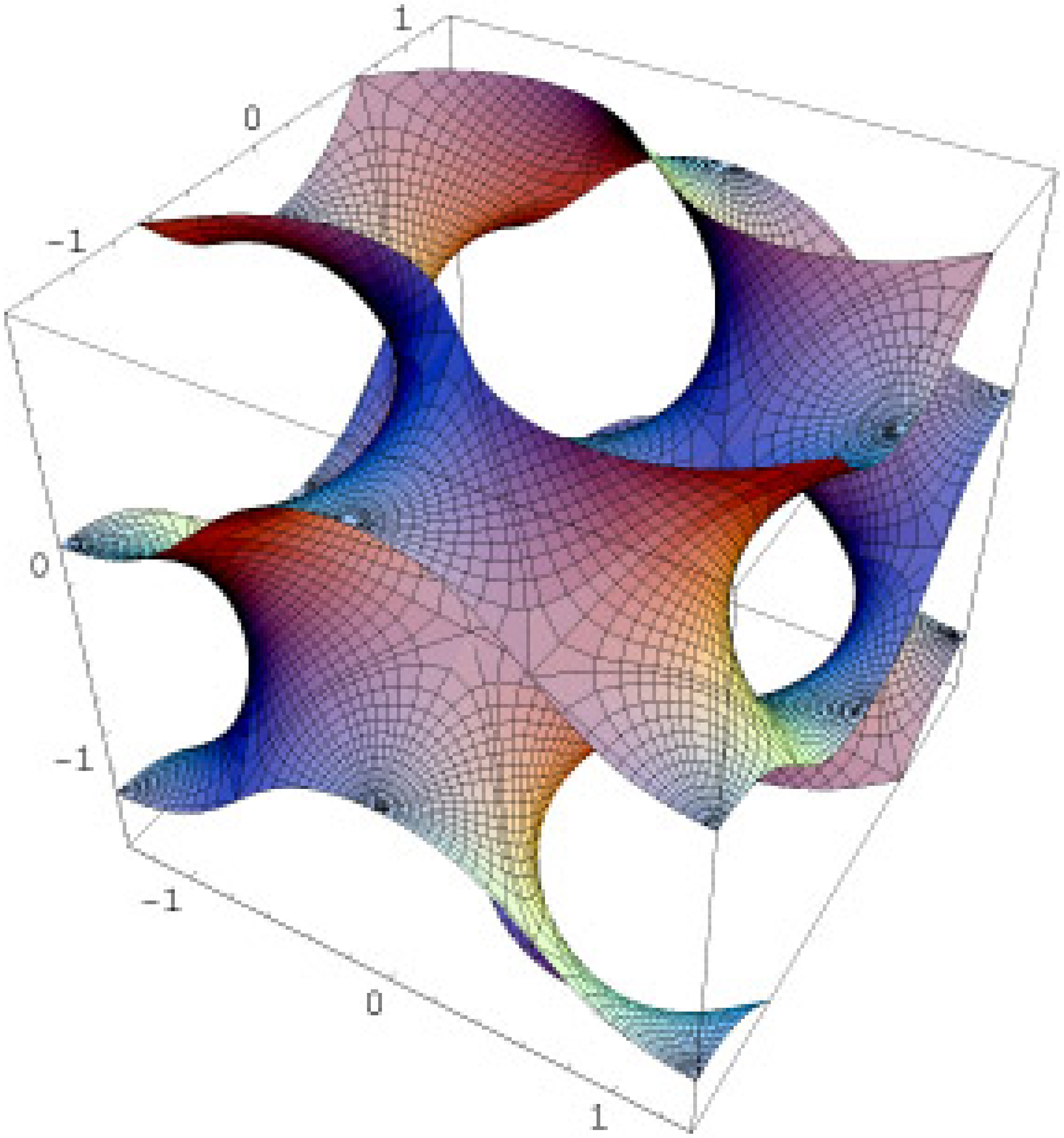}
\leavevmode\includegraphics[width=40mm]{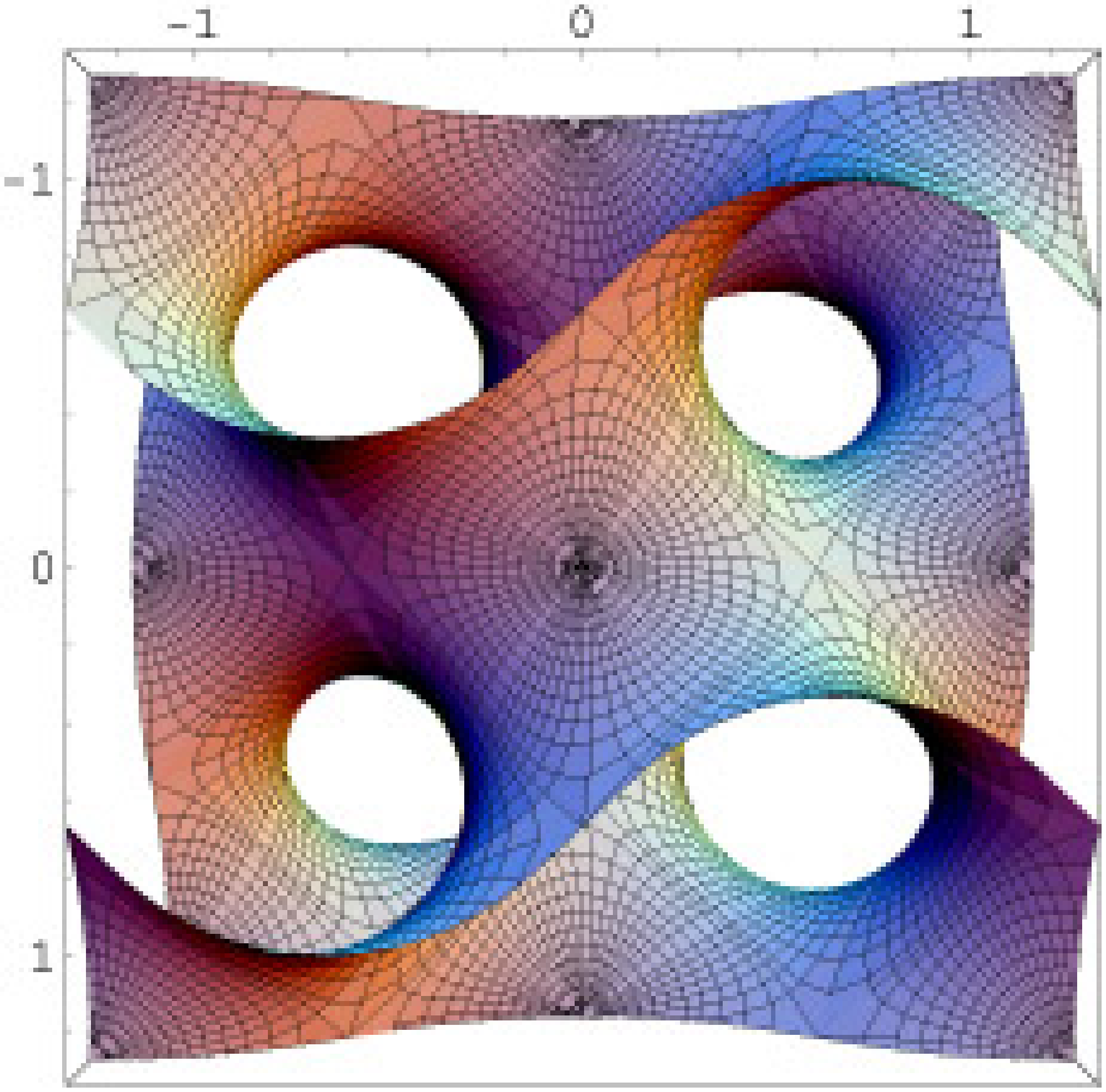}
\end{center}
\caption{
A cubic unit cell of G-surface in birds-eye(a) 
and top(b) views.
}
\label{fig_cell}
\end{figure}

Figure \ref{fig_patch} shows a primitive patch of G-surface,
which corresponds to eqn.(\ref{eqn:Weiermap}) 
with $0 < \theta < \pi$, $-\pi/4<\phi<\pi/4$ 
when we stereographically map 
$(u,v)$ to a unit sphere $(\theta, \phi)$ 
with $w=u+iv\equiv \cot(\theta/2) {\rm e}^{i\phi}$.  
The full surface is depicted in Fig. \ref{fig_cell} 
for its cubic unit cell (containing two bcc cells).   
The surface has 90$^\circ$ helical symmetry axes
along $x$, $y$ and $z$ directions, respectively, 
while the chirality
(right- or left-handedness) is opposite across neighboring helices.  
This is how the surface is a network of helical tubes along $x,y,z$ 
connected into a single labyrinth.  

While the helical structure is unique to the G-surface, 
it shares with P- and D- surfaces the property called the balance surface 
(or {\it the in-out symmetry}), which is 
defined as the curved surface that divides the three-dimensional 
space into 
two, congruent labyrinths.  For the G-surface the divided 
spaces are mirror inversions of each other.
The surface itself is not chiral, being symmetric
with respect to flat points (the center of the primitive patch
in Fig. \ref{fig_patch}) which corresponds to the 
navels. 
It has been shown\cite{Fischer} that a balance surface is characterized 
by a pair of space groups ($G, H$), where $G$ maps one side of the 
surface to itself or the other side (and one labyrinth to itself or 
the other labyrinth), while $H$, a subgroup of $G$, maps 
each side to itself (and each labyrinth to itself).  
For G-surface, $(G,H)=(Ia\bar{3}d,I4_132)$.

Schr\"{o}dinger's equation on a 
curved surface takes different forms between the following two cases; 
one is to consider electrons bound 
to a thin, curved slab of thickness $d$, where
the limit $d\rightarrow 0$ is taken\cite{Nagaoka}, while the 
other is to ignore the perpendicular degree of freedom from the outset.   
We adopt the former as a physical approach, 
for which the equation reads\cite{Koshino}
    \begin{eqnarray}
      &&\left[  - {\hbar^2 \over 2m}
        {1 \over \sqrt{g}}
        {\partial \over \partial q^i} \sqrt{g}\ g^{ij}
        {\partial \over \partial q^j}
        - {\hbar^2 \over 8m}\left( \kappa_1 - \kappa_2 \right)^2
      \right] \psi(q^1, q^2) \nonumber \\ &&\hspace{45mm} = E\ \psi(q^1, q^2),
      \label{eqn:SchResult2}
    \end{eqnarray}
where $(q^1,q^2)\equiv (u,v)$, $g_{ij}$ the metric tensor with 
summations over repeated indices assumed, 
and $\kappa_1,\kappa_2$ the local principal curvatures 
(with $\kappa_1+\kappa_2=0$ everywhere for a minimal surface 
by definition).  
An effect of the curvature of the surface 
appears as a curvature potential, 
$-(\hbar^2/8m)(\kappa_1 - \kappa_2)^2$, which 
has its minima at navels.

As shown in Ref.\cite{Koshino}, 
the Weierstrass-Enneper representation transforms 
Schr\"{o}dinger's equation into a neat form of
\begin{eqnarray}
      -{ (1 - \cos\theta)^4 \over |F|^2 }
      \bigg(
              {\partial^2 \over \partial\theta^2}
              + \cot\theta{\partial \over \partial\theta}
               + {1 \over \sin^2 {\theta}}{\partial^2 \over
 \partial\phi^2} +  1\bigg) \psi  \nonumber\\
 = \varepsilon \ \psi.
      \label{eqn:SchOnP}
    \end{eqnarray}
Since the Bonnet transformation preserves 
the metric tensor as well as the Gaussian curvature, the
surfaces connected by a Bonnet transformation obey the identical 
Schr\"{o}dinger's equation, as seen in 
eqn.(\ref{eqn:SchOnP}) where $F$ only
enters as $|F|$.  However, this only applies to a unit patch, so 
that the band 
structures are different between P,D, and G, 
since the way in which unit patches are
connected is different.\cite{Koshino}

\begin{figure}
\begin{center}
\leavevmode\includegraphics[width=60mm]{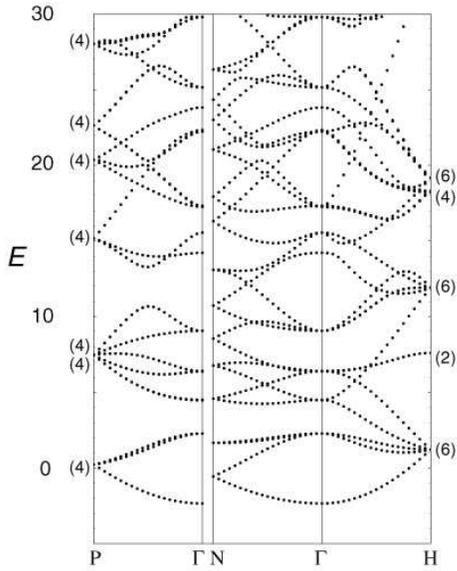}
\end{center}
\caption{
  Band structure of the G-surface displayed on the 
  Brillouin zone for the bcc unit cell.  
  The numbers in the parentheses represent the degeneracy 
  of the band sticking.  
}
\label{fig_band}
\end{figure}

\begin{figure}
\begin{center}
\leavevmode\includegraphics[width=70mm]{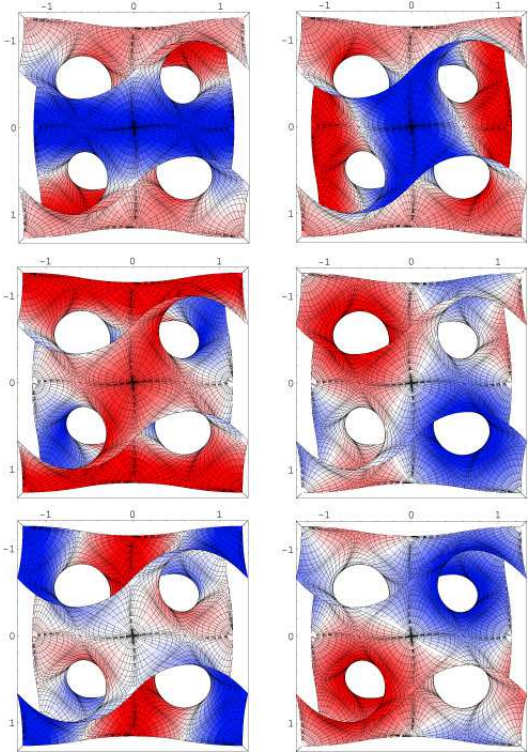}
\end{center}
\caption{
The wave functions in the unit cell of G-surface
for the six-fold states
at the lowest level in H-point (in Fig. \ref{fig_band}), 
with positive (negative) amplitudes color coded in red (blue).
}
\label{fig_wave}
\end{figure}

So we have solved the equation for the appropriate 
connection between the patches to obtain the band structure for G-surface 
as shown in Fig. \ref{fig_band}.
We can immediately notice that the bands are 
multiply degenerate at some symmetric points 
on the Brillouin zone boundary:
six-fold degenerate at H-point, four-fold at P, etc, 
which is a phenomenon not seen in P and D-surfaces.  
The wave functions for six-fold states at H
are displayed in Fig. \ref{fig_wave}. 
The space group for the G-surface ($Ia\bar{3}d$) 
is non-symmorphic, i.e., has the screw axes 
or glide planes, and this should cause the degeneracy 
at the zone boundary, generally 
known as `the band sticking together'\cite{Heine} 
in e.g. crystals of Se and Te with helical atomic arrangements.  
We shall elaborate this point below.

\begin{figure}
\begin{center}
\leavevmode\includegraphics[width=60mm]{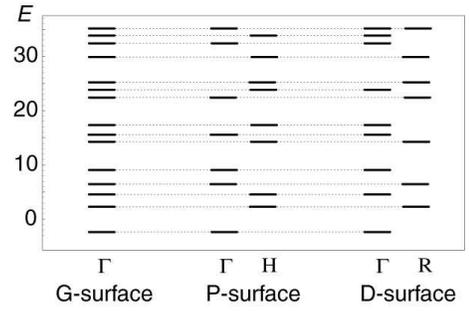}
\end{center}
\caption{
Coincidence of the band energies among G-, P- and D-surfaces.
}
\label{fig_band_GPD}
\end{figure}

We can note in passing the following: We have pointed out 
in the previous work\cite{Koshino} that 
Bonnet-connected P- and D-surfaces, despite the different band structures, 
have a common set of band energies 
at special $k$-points (Brillouin zone corners, edges, or face-centers). 
This occurs when a wave function on a unit patch 
can be continued as a ``tiling'' of patches on both surfaces, where 
the simplest case is the ground state at $\Gamma$ point.  
In G-surface, this occurs at (and only at; a 
peculiarity of the spiral structure) 
$\Gamma$ at which the band energies accommodate 
those of P-surface at $\Gamma$ and H points 
and D-surface at $\Gamma$ and R as displayed in Fig.\ref{fig_band_GPD}. 

\begin{figure}
\begin{center}
\leavevmode\includegraphics[width=77mm]{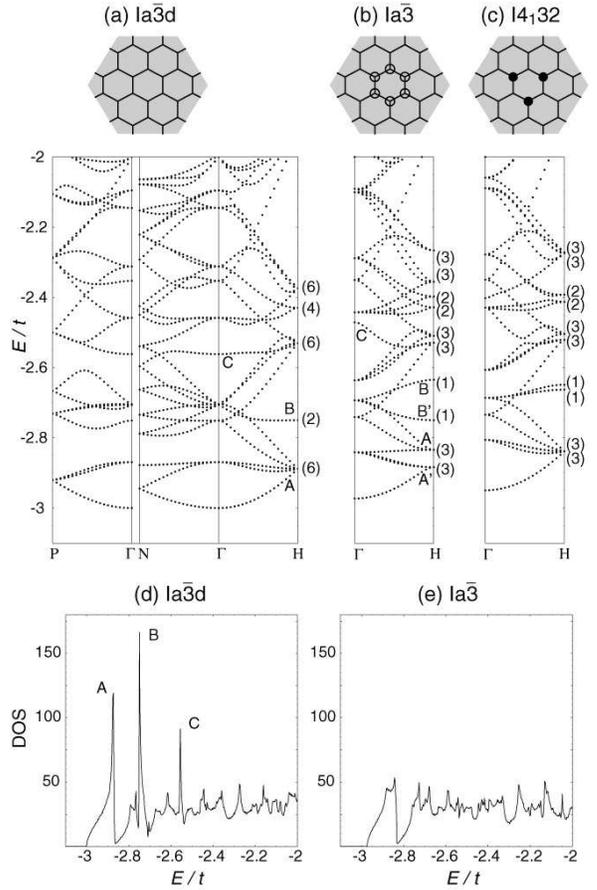}
\end{center}
\caption{
(a) Low-energy band structure for the graphitic G-sponge,
to be compared with Fig. \ref{fig_band},
constructed from a graphite fragment shown at the top.
We take here the hopping parameter $t$ as the unit of energy 
and the tight-binding band center as the origin.  
(b) The band structure when we introduce a potential that breaks 
the helical symmetry, or (c) the inversion symmetry.
Extra potential introduced to degrade the symmetry is shown at the top,
where an open circle represents the potential for every other patch,
while filled one for all the patches.
The panels (d) and (e) show 
the density of states for the systems (a) and (b), respectively, 
in units of $1/(Vt)$ ($V$: unit cell volume).
}
\label{fig_honeycomb}
\end{figure}


A question from the materials science point of view is: 
can we realize the G-surface in some materials?  
As mentioned above, there are a few classes of 
materials that possess the same space group as the 
G-surface.  One is a class of clathrate compounds 
of group-IV elements (e.g., Ge), where Ge$_{20}$ clusters that 
include another element (e.g., Ba) are 
stacked in a triply-helical fashion.  
A conceptually simpler system would be a three-dimensional 
labyrinth of graphite sheets that forms a triply 
periodic surface.  Fujita and coworkers have considered 
such systems, and called them ``graphitic sponges''\cite{Fujita}.  
The negative-curvature fullerene
(or C$_{60}$ zeolite)\cite{MacTerr} 
is indeed a realization of the P-surface if we smear out atoms into a 
surface.\cite{Lenosky}   
The band structure of atomic networks such as the C$_{60}$-zeolite 
is expected to basically reflect the properties of an electron 
on the curved surface, 
as far as the effective-mass formalism is applicable 
and effects of the odd-membered atomic rings are neglected.  
This has in fact been shown for P- and D-surfaces\cite{KoshinoEP2DS}.  
Fabrication of graphitic sponges has been experimentally attempted 
with a zeolite as a
template\cite{kyotani}, so we should end up with a G-surface 
sponge if we use a G-structured zeolite (MCM) as the template.

The electronic structures of the negative curvature carbon networks,
including G-surface-like structures, 
were investigated with a first
principles method in Ref.\cite{Huang}.  However, 
some of the G-surface-like models (E and O in that paper) 
contain seven-membered 
rings, which degrade the symmetry below that of the G-surface, 
and the band degeneracy was not discussed, either.  
Here we examine the band structure of the 
graphite sponge that shares the global topology 
and the symmetry with the G-surface 
to investigate how the degeneracy would be lifted
when the symmetry is degraded.
We constructed the network by arraying the unit patches 
displayed in Fig. \ref{fig_honeycomb}(a)
(corresponding to Fig. \ref{fig_patch}), where the resultant 
structure contains no odd-membered rings, 
and the band structure is calculated in the one-band, 
tight-binding model with the hopping parameter $t$.

The result in Fig. \ref{fig_honeycomb}(b) shows that
the bands in the low-energy region have a one-to-one correspondence with 
those for the G-surface, as expected, 
so the electronic structure there is basically 
determined by the structure of the surface on which the atoms reside.
Then we have degraded each of the two symmetries (helical 
and inversion), by introducing an extra potential.  
We first destroy the helical symmetry in Fig.\ref{fig_honeycomb}(b), 
with the space group $Ia\bar{3}d$ reducing to $Ia\bar{3}$, 
with the extra potential introduced on every other patch.  
We next destroy the inversion in Fig.\ref{fig_honeycomb}(c) 
(with the space group reducing to $I4_132$) 
with the extra potential on triangular sites around the 
center of every patch.
The amplitude of the extra potential energy is taken to be $0.3t$ 
or $0.5t$ in (b), (c), respectively, 
which are small enough for the correspondence among the bands 
to be retained. 
The band structures show that
the multifold stickings split into lower degeneracies,
although some degeneracies in (c) (with the inversion symmetry broken)
are lifted only slightly. This endorses that 
both the helical and inversion symmetries 
are essential for the band sticking.
The panels (d) and (e) display 
the density of states (DOS) before(d) and after(e) the helical symmetry 
is broken.  
We recognize that the first and second prominent peaks 
(labeled as A and B, 
corresponding respectively to six-fold and two-fold stickings in (a))
split with significantly reduced peak heights after the symmetry 
is degraded.  The effect is more noticeable for the peak B, 
which is due to the splitting of nearly flat bands.
The density of states for the broken inversion symmetry 
(not shown) exhibits a similar behavior.  So the band sticking due to
the G-surface space group 
can generally give rise to prominent peaks in the density of states.  
The third peak (C) is also suppressed, 
but this is caused by a change in the band curvature.

Let us finally mention a possible relevance to real materials.  
A recent band calculation for the clathrate Ba$_6$Ge$_{25}$, 
which shares the space group with G-surface, 
shows that the DOS peak at the Fermi level 
becomes split when the structural symmetry is reduced, 
which they suggest causes a band Jahn-Teller instability, 
and should be responsible for an 
experimentally observed phase transition\cite{Zerec}.
Our result suggests that materials with the G-surface  
symmetry generally have a potential for such instabilities 
and corresponding phase transitions.

We wish to thank Shoji Yamanaka and Yasuo Nozue 
for illuminating discussions 
in the early stage of the study.

\end{document}